\documentstyle[preprint,aps]{revtex}
\begin{document}
\draft
\title{
Nuclear Mass Dependence of Chaotic Dynamics \\
in Ginocchio Model}
\author{
Naotaka Yoshinaga
}
\address{
Department of Physics, College of Liberal Arts, Saitama University
Urawa-shi, Saitama 338, Japan}
\author{
Nobuaki Yoshida
}
\address{
Goto Laboratory, Institute of Physical and Chemical Research (RIKEN)
Wako-shi, Saitama 351-01, Japan}
\author{
Takaomi Shigehara
}
\address{
Computer Centre, University of Tokyo, Yayoi, 
Bunkyo-ku, Tokyo 113, Japan 
}
\author{
Taksu Cheon
}
\address{
Department of Physics, Hosei University, Fujimi, 
Chiyoda-ku, Tokyo 102, Japan
}
\date{April 5, 1993}
\maketitle
\begin{abstract}
The chaotic dynamics in nuclear collective motion is studied in the 
framework of a schematic shell model which has only monopole and 
quadrupole degrees of freedom. The model is shown to reproduce the 
experimentally observed global trend toward less chaotic motion in 
heavier nuclei.  The relation between current approach and the earlier 
studies with bosonic models is discussed.
\end{abstract}
%
%
%
%
\narrowtext
Several recent studies have revealed an intriguing interplay of chaotic 
and regular motion in the low-lying collective states of the medium-
heavy nuclei [1-3].  A close relationship between the breaking of 
dynamical symmetry and the emergence of chaotic motion has been 
displayed.  All the realistic studies up to now have employed the 
interacting boson model of Arima and Iachello [4] which is well suited 
for the study of the dynamical symmetry with its group theoretical 
formulation.  It should not be forgotten, however, that the bosonic 
models are approximations to the underlying fermionic models, and 
that despite their success they have their intrinsic limitations.  One of 
the places where bosonic models fail is the physics of higher 
excitations where the explicit treatment of the broken fermion pairs 
is called for [5].  Even within the low-lying regime, the effects of the 
underlying fermionic degrees of freedom might be felt in such subtle 
question as chaos and order.  Moreover, there are certain problems 
related to the global change of the nuclear characteristics.  Usually 
they are better approached by the microscopic fermionic models 
because these problems are often related to the shell effect and the 
Pauli-blocking effect.

	One of such global-change problems has recently been brought 
into light by Shriner, Mitchell and von Egidy [6] who have compiled 
the spectral statistics of nuclear levels 
ranging from $^{24}$Na to $^{244}$Am.  
The fluctuation properties of the nuclear levels are found to vary in 
different mass region; generally, the level spacing distribution is 
"chaotic" or Wigner-like among lighter nuclei, and tends to become 
"regular" or Poisson-like among heavier nuclei.  Since large portion of 
the nuclear levels are of "rotational" nature, it is now clear that we 
have to depart from the notion that the rotational levels are 
necessarily regular as suggested by the SU(3) limit of interacting boson 
model.  In fact, no dynamical symmetry is known in the deformed 
limit of the realistic shell model.  We face the problem of examining 
where and how the regular spectra emerge from the complexity of the 
fermionic shell model.

	There is, however, a long way from the bosonic models to the 
full-fledged shell model.  Several attempts have been made to bridge 
these two models through some justifiable truncation.  One of the 
more successful among them is the model developed by Ginocchio [7], 
in which the fermions are allowed to be coupled to monopole and 
quadrupole pairs only.  This model has an algebraic structure which 
benefits both the physical intuition and the ease of computation, a 
feature reminiscent of the interacting boson model.  Despite its 
schematic nature, the Ginocchio model has proven to be a surprisingly 
good phenomenology for samarium isotopes [8].  The Ginocchio model 
with its simplicity and flexibility is an ideal tool for the study of the 
fermionic effects on the chaotic nuclear dynamics.

	In this Report, we examine the spectral statistics of the 
Ginocchio model.  We first focus on the transition between vibrational 
and rotational dynamics.  The results are compared to the earlier 
studies based on the interacting boson model.  We then proceed to the 
problem of global change of the degree of chaos.  It is shown that the 
experimental trend toward the regular motion in heavier nuclei 
naturally emerges from the Ginocchio model with the different shell 
degeneracy.
	We start from a single j-shell model in which a nucleus is 
described as a system of fermions with spin $i$ filling a shell specified 
by its orbital angular momentum ${\vec k}$ and total angular momentum
${\vec j} = {\vec k} + {\vec i}$.  
Ginocchio has conceived a model [7] where the intrinsic spin of the 
fermion is given by $i$ = 3/2 in place of the nucleon's one-half.  We refer 
to this fictitious spin as the pseudo spin.  In the L-S coupling scheme, 
the two-particle state is specified by its total orbital and spin angular 
momenta ${\vec K} = {\vec k_1}+{\vec k_2}$ and 
${\vec I} = {\vec i_1} + {\vec i_2}$.  
The identity of the two fermions prohibits the odd 
values for the total spin, namely only $I$ = 0 and $I$ = 2 are allowed.  
The unique feature of the Ginocchio model is in the fact that the operators 
made of two fermions coupled to $K$ = 0 form a closed algebra SO(8) 
whose 28 generators are given by
\begin{mathletters}
\label{1}
\begin{eqnarray}
\label{1a}
S^\dagger \ =\sqrt {2k+1}\ [\ b_{ki}^\dagger \ b_{ki}^\dagger \ ]^{(0,0)}
\end{eqnarray}
\begin{eqnarray}
\label{1b}
S\ =\sqrt {2k+1}\ [\ \tilde b_{ki}\ \tilde b_{ki}\ ]^{(0,0)}
\end{eqnarray}
\begin{eqnarray}
\label{1c}
D^\dagger =\sqrt {2k+1}\ [b_{ki}^\dagger \ b_{ki}^\dagger
]^{(0,2)} \end{eqnarray} %
\begin{eqnarray}
\label{1d}
\tilde D =\sqrt {2k+1}\ [\tilde b_{ki} \ \tilde b_{ki}]^{(0,2)}
\end{eqnarray}
\begin{eqnarray}
\label{1e}
R^{(L)} =\sqrt {2k+1}\ [b_{ki}^\dagger\  \tilde b_{ki}]^{(0,L)}
\ \ \ L = 0, 1, 2, 3
\end{eqnarray}
\end{mathletters}
where $b_{ki}^\dagger $
 and $b_{ki}$ are the fermion creation and annihilation operators.  
The notation $(0,L)$ means that the pseudo orbital angular momenta are 
coupled to 0 and the pseudo spins to L.  The number operator of 
fermions n is given by $n = 2 R^{(0)}$.  We also define $N = n/2$ which 
represents the number of pairs for later use.

	The SO(8) symmetry of this model has three dynamical sub-
symmetries SO(5) $\times$ SU(2), SO(6) and SO(7), all sharing the common 
subgroup SO(5) in addition to the rotational symmetry SO(3), namely,
\begin{eqnarray}
\label{2}
SO(8)\supset \left\{ {\matrix{{SO(5)\times SU(2)\ }\cr
{SO(6)}\cr
{SO(7)}\cr
}} \right\}\supset SO(5)\supset SO(3)
\end{eqnarray}
The common subgroup SO(5) poses a problem in the actual application 
of the model, since no such universal symmetry is known among real 
nuclei.  This difficulty has been solved in the extended model 
developed by Arima, Ginocchio and Yoshida [8] where  protons and 
neutrons are distinguished.  In our current work, we adopt a simplified 
version of this extended model, whose hamiltonian is given by
\begin{eqnarray}
\label{3}
H\ =-g_\pi S_\pi ^\dagger S_\pi 
-g_\nu S_\nu ^\dagger S_\nu 
-\kappa Q_\nu ^{(2)}\cdot Q_\pi ^{(2)}
\end{eqnarray} %
where quadrupole operator $Q^{(2)}$ is defined by
\begin{eqnarray}
\label{4}
Q^{(2)}=R^{(2)}-{\textstyle{{\sqrt 7} \over {2\Omega }}}
[\ D^\dagger \ \ \tilde D\ ]^{(2)}
\end{eqnarray}
The subscripts $\pi$ and $\nu$ in eq. (3) refer to the proton and 
neutron respectively.  The first two terms in eq. (3) represent the 
paring interactions between like particles, while the last term 
represents the quadrupole interaction between different particles. 
Thus this hamiltonian has the essential feature of the "Paring plus 
QQ" model. The $QQ$ interaction explicitly breaks the unwanted sub-
symmetry SO(5).  The shell degeneracies $\Omega_\pi$ and 
$\Omega_\nu$ appearing in eq. (4) are defined in terms of the values 
of orbital angular momenta $k_\pi$ and $k_\nu$ as $\Omega_\pi$ = 2 
(2$k_\pi$+1) and $\Omega_\nu$ = 2 (2$k_\nu$+1).  With this 
definition the shell degeneracy stands for half of the maximum 
number of fermions in each shell.  It is known that the Ginocchio 
model approaches to the interacting boson model when $\Omega_\pi$ 
and $\Omega_\nu$ approach infinity [7,9].  In physical term, the 
collectivity realized at $\Omega = \infty$ is hindered by the Pauli-
blocking effect when $\Omega$ is small.  Without losing the essential 
feature, we simplify the hamiltonian by introducing a single 
parameter x and putting the constraints among the three parameters 
$g_\pi$, $g_\nu$ and $\kappa$ as
\begin{eqnarray}
\label{5}
g_\pi =g_\nu =1-x, \ \ \kappa =5\ x
\end{eqnarray}
When the parameter $x$ is varied from 0 to 1, we can simulate nuclei 
from pairing-vibrational to rotational limits.  The factor 5 in the 
second of eq. (5) is chosen just for convenience, and it can be re-scaled 
to any value without changing the physical content.  The SO(5) 
dynamical symmetry exists in the pairing-vibrational limit.  On the 
contrary, no dynamical symmetry is present in the rotational limit of 
the Ginocchio model, except in the limiting case of  $\Omega_\pi  = 
\Omega_\nu = \infty$   where the rotational limit becomes the SU(3) 
limit of the interacting boson model [7,9].  

	The exact relation between the level statistics of a quantum
system and the dynamics of its classical counterpart is still an open
problem in the quantum chaology [10,11].  There exist, however, a fair
amount of theoretical and numerical researches supporting the standard
assumption of quantum level statistics, the correspondence between the
order to chaos transition in the classical system and the Poisson to
Wigner statistics in the nearest neighbor level spacing distribution
in the quantum counterpart [10-16].  The example of the interacting
boson model indicates that this association might be taken seriously
to a certain semi-quantitative level [2,3].  On this premise, we
examine the nearest level spacing distribution $P(s)$ to probe the
degree of chaos in the nuclear levels.

	We look at the change in the level statistical of the
Ginocchio model along the vibrational and rotational limits as we vary
the value of x from zero to one.  In Fig. (1), the nearest level
spacing distributions $P(s)$ at $x$ = 0.1, 0.2, 0.4 and 1 are displayed.
Throughout, the shell degeneracies and the numbers of the pairs are
kept to be $\Omega_\pi$ = 11, $\Omega_\nu$ = 22, and $N_\pi$ = 5,
$N_\nu$ = 4, which corresponds to the $^{150}$Nd nucleus.  One
observes that the $P(s)$ starts out as Poisson-like around the pairing
limit, then becomes Wigner-like at intermediate region $x$ = 0.4.
Further to the rotational limit, it stays Wigner-like up to $x$ = 1.
The result in Fig. (1) can be interpreted as the dynamics of actual
rotational nuclei being more chaotic than the previous studies based
on the interacting boson model indicate.  In hindsight, this is a
natural consequence of the absence of dynamical symmetry in the
rotational limit of the Ginocchio model.

	The Fig. (2) shows how large $\Omega$ one needs in order to
reach the SU(3) limit of the interacting boson model in terms of the
level statistics.  Here, $x$ is fixed to be one and $\Omega = \Omega_\pi
= \Omega_\nu$ is varied as 10, 20, 40, 80.  $N_\pi$ and $N_\nu$ are
again fixed to be 5 and 4.  One observes that the Poisson statistics
as in the boson limit is certainly recovered, but only at very large
value of $\Omega$.

	The result shown in Fig. (2) also has an interesting
implication to the global trend across the different nuclear mass
region.  Within the framework of single-j model, larger mass number
corresponds to the lager value of $\Omega$ for the valence shell.
Therefore, the transition to Poisson-like statistics in Fig. (2) seems
to offer a natural explanation for the experimentally observed global
shift toward the regularity in the spectra of heavier nuclei [6].  In
order to make a simple estimate, we assume that most nuclei are
deformed, and most levels are rotational.  We also assume the spectral
statistics of the whole range of nuclei in a major shell is
represented by a single typical nucleus.  These assumptions may appear
quite drastic, but we argue that they still keep the underlying
physics intact.  For the first point, it can be pointed out that even
spherical nuclei display rotational bands in their excited states
whose number tends to dominate over the sparsely found single particle
excitations.  For the second point, we may recall that a fixed
parameter set has successfully described the entire samarium isotopes
[8].  Also, we have checked in the actual calculation that the results
are insensitive to the modest variation of the parameters.

        In Table (1), the Brody parameters [17] of the level spacing
statistics in the various nuclear mass regions are tabulated.  Four
columns correspond to ( $\Omega_\pi , \Omega_\nu$ ) = (11, 11), (16,
16), (16, 22) and (22, 29), each roughly representing nuclei of the
mass regions of 50 $<$ A $<$ 100, 100 $<$ A $<$ 150, 150 $<$ A $<$ 210
and 210 $<$ A, respectively.  In the second line, theoretical
estimates of the Brody parameter $\beta$ are shown.  They are
calculated from the second moment of P(s) as in Ref. [3].  The
parameter x is fixed to be one in accordance with our assumption. The
numbers of the fermion pairs are set to be $N_\pi$ = 5 and $N_\nu$ = 4
as a typical example.  The number of the levels included in the
calculation is 900, of which 80 are of $L$ = 0 levels, 250 of $L$ = 2
levels, 250 of $L$ = 3 and 320 of $L$ = 4.  In the forth line of the
Table (1), the "experimental" values of the Brody parameter for
even-even nuclei taken from the paper of Shriner {\em et al.}  [6] are
tabulated. In the fifth line we show the same parameter for all
nuclei. The agreement is as good as one can expect from our simplistic
estimate and marginal statistics.  In fact, we are unable to give, for
example, the statistical error of the theoretical value of Brody
parameter, since the model hamiltonian employed has very little
apparent resemblance to the realistic hamiltonian.  However, we stress
that the basic properties of the shell model hamiltonian, namely the
dominance of pairing and the quadrupole degrees of freedom and the
associated dynamical symmetry are nicely captured in the Ginocchio
model. This is evidenced both in its bridging role between shell model
and the phenomenologically successful interacting boson model [7], and
also in its own success as a phenomenology for the samarium isotopes
with a simple hamiltonian very similar to the one we have employed in
this work [10].  Thus we assert that the experimentally observed trend
in the nuclear level statistics is physically understood within a
schematic single j-shell model: namely, the chaotic spectrum in
lighter nuclei is the result of the incomplete realization of
dynamical symmetry due to the Pauli-blocking effect.  Concerning the
spin effect reported by Abul-Magd and Weidenm{\" u}ller [18], we are
unable to see the difference in the level statistics of $L$ = 0, 3
states and $L$ = 2, 4 states in the current model calculations.  This
and other experimental characteristics will be the subjects of the
forthcoming publication along with the more detailed description of
our calculations.

	In summary, we have examined the spectral statistics of  a 
model hamiltonian in the Ginocchio model in order to probe the 
chaotic dynamics in the nuclear low-lying states across broad nuclear 
mass region.  The results show several novel features which were not 
noticed in previous studies with phenomenological bosonic models.

	We acknowledge our gratitude to Prof. A. Arima for calling our
attention to the Ginocchio model.  Special thanks are due to
Dr. H. Nakada for his critical reading and constructive comments of
the manuscript.  Thanks are also due to Dr. Mizusaki, Prof. T. Otsuka,
and Mr. M. S. Bae for useful discussions.  The numerical work has been
performed on the Hitachi M880 at Computer Centre, University of Tokyo,
on VAX 6000/440 of Meson Science Laboratory, University of Tokyo, and
on SPARCStation 2 at Hosei University Department of Physics.

\begin{figure}
\caption{
The level spacing distribution $P(s)$ of the Ginocchio model with the 
various values of $x$, with Wigner (solid lines) and Poisson (dotted lines) 
formulae.
}
\end{figure}
\begin{figure}
\caption{ 
$P(s)$ for various $\Omega$ at the rotational limit $x$ = 1.  
\ \ \ \ \ \ \ \ \ \ \ \ \ \ \ \ \ \ \ \ \ \ \ \ \ \ \ \ \
\ \ \ \ \ \ \ \ \ \ \ \ \ \ \ \ \ \ \ \ \ \ \ \ \ \ \ \ \
}
\end{figure}
\begin{table}
\caption
{ The Brody parameter $\beta$ at various mass region.  Here $\beta$ =
0 corresponds to the Poisson (or regular) statistics, and $\beta$ = 1
to the Wigner (or chaotic) statistics.  The $\beta$(calc) means
calculated Brody parameter in the present model. The $\beta$(exp)E
indicates the experimentally compiled Brody parameter for even-even
nuclei, while $\beta$(exp)A , for all nuclei }
\begin{tabular}{|c|c|c|c|r|} \hline
\multicolumn{1}{|c} {$\Omega_\pi/\Omega_\nu$} &
\multicolumn{1}{c|} {$\beta$(calc)} &
\multicolumn{1}{c} {A} &
\multicolumn{1}{c} {$\beta$(exp)E} &
\multicolumn{1}{c|} {$\beta$(exp)A} \\ \hline
 11/11 & 0.74 & 50--100 &     --      & 0.88$\pm$41 \\ \hline 
 16/16 & 0.60 &100--150 & 0.62$\pm$16 & 0.55$\pm$11 \\ \hline 
 22/16 & 0.51 &150--180,180-210& 0.26$\pm$11, 0.30$\pm$18 & 
0.33$\pm$07, 0.43$\pm$17 \\ \hline 
 29/22 & 0.37 & 210--   & 0.27$\pm$32 & 0.24$\pm$10 \\ \hline    
\end{tabular} \end{table} 
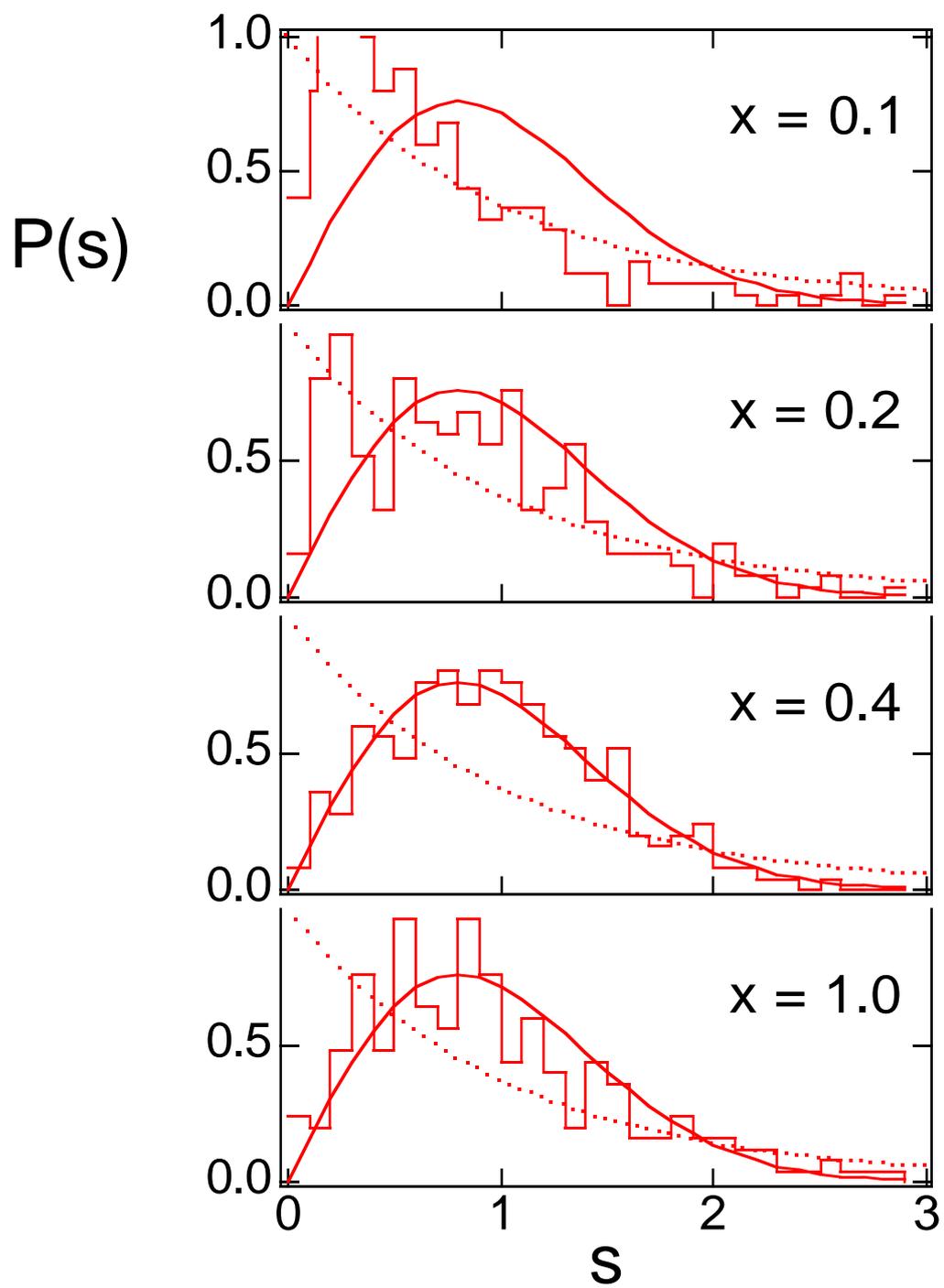
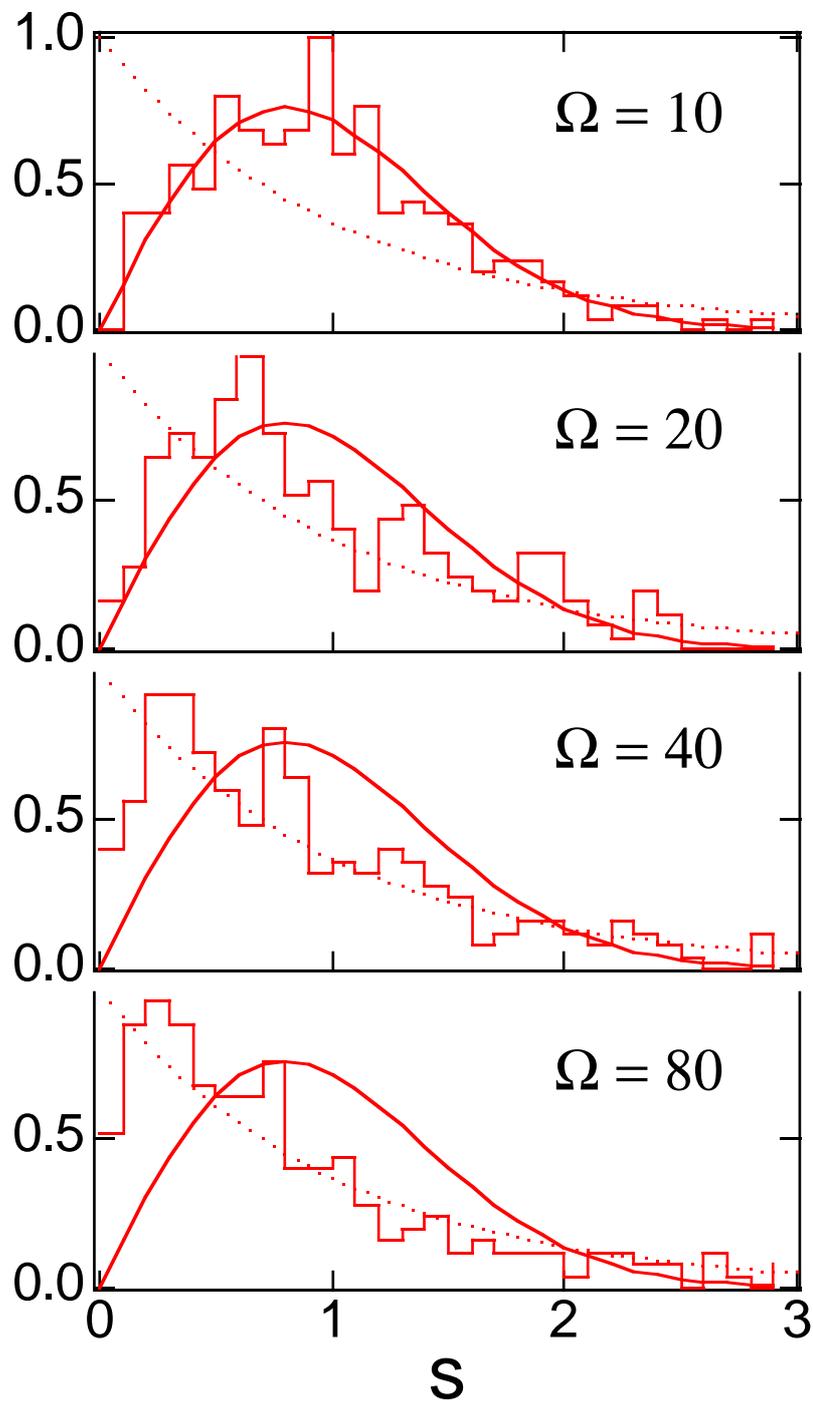
\end{document}